\documentclass[a4paper,12pt,dvipsnames,usenames]{article}
\pdfoutput=1

\newcommand{\vecU}{\ensuremath{\overrightarrow{U^2}}}

\newcommand{\fship}{\texttt{FairShip}\xspace}

\newcommand{\decay}{\text{\rm decay}}

\usepackage{xspace}
\usepackage[]{graphicx}
\graphicspath{{./figs/}}

\usepackage{jheppub} 

\usepackage{overpic}
\newcommand{\SHiPOverlay}[5]{
  \begin{overpic}[#1]{#2}
      \put(#3,#4){\includegraphics[#5]{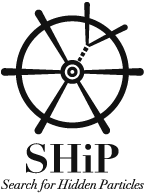}}
    \end{overpic}
}
\usepackage{amsmath,amssymb,bm}
\usepackage{array}
\usepackage{color,xcolor,colortbl}
\usepackage[utf8]{inputenc}
\usepackage[english]{babel} 
\usepackage{lineno}
\usepackage{paralist}
\usepackage[section]{placeins}
\usepackage{slashed}
\usepackage[textsize=scriptsize,color=ForestGreen!05,linecolor=gray]{todonotes}
\usepackage{units}
\usepackage{xspace}
\usepackage{longtable}

\newcommand{\event}{\text{\rm events}}
\newcommand{\pr}{\text{\rm prod}}

\newcommand{\BR}{\mathop{\text{BR}}\nolimits}
\newcommand{\Jpsi}{{\ensuremath{J/\psi}}}

 \usepackage[colorlinks=true,citecolor=blue]{hyperref}

\begin{document}

\title{Sensitivity of the SHiP experiment to Heavy Neutral Leptons} \collaborationImg{\rightline{\includegraphics[scale=0.5]{SHIP_Logo1-146x195px}}}
\collaboration{The SHiP collaboration}

--------------------------------------------------------------------------------
\author[1]{C.~Ahdida}
\author[2,a]{R.~Albanese}
\author[2]{A.~Alexandrov}
\author[3]{A.~Anokhina}
\author[4]{S.~Aoki}
\author[1]{G.~Arduini}
\author[5]{E.~Atkin}
\author[6]{N.~Azorskiy}
\author[1]{F.~Baaltasar~Dos~Santos}
\author[7]{J.J.~Back}
% ...

\affiliation[1]{European Organization for Nuclear Research (CERN), Geneva, Switzerland}
\affiliation[2]{Sezione INFN di Napoli, Napoli, Italy}
\affiliation[3]{Skobeltsyn Institute of Nuclear Physics of Moscow State University (SINP MSU), Moscow, Russia}
\affiliation[4]{Kobe University, Kobe, Japan}
\affiliation[5]{National Research Nuclear University (MEPhI), Moscow, Russia}
\affiliation[6]{Joint Institute for Nuclear Research (JINR), Dubna, Russia}
\affiliation[7]{University of Warwick, Warwick, United Kingdom}
% ...
\affiliation[{a}]{Universit\`{a} di Bari, Bari, Italy}
% ...

\abstract{
Heavy Neutral Leptons (HNLs) are hypothetical particles predicted by many extensions of the Standard Model.
These particles can, among other things, explain the origin of neutrino masses,  generate the observed matter-antimatter asymmetry in the Universe and provide a dark matter candidate.

The SHiP experiment will be able to search for HNLs produced in decays of heavy mesons and travelling distances ranging between $\mathcal{O}(50\text{ m})$ and tens of kilometers before decaying.
We present the sensitivity of the SHiP experiment to
a number of HNL's benchmark models and provide a way to calculate the SHiP's sensitivity to HNLs for arbitrary patterns of flavour mixings.
The corresponding tools and data files are also made publicly available. }

\arxivnumber{1811.00930}

\maketitle
\flushbottom
\section{The SHiP experiment and Heavy Neutral Leptons}

\paragraph{The SHiP experiment.}
The Search for Hidden Particles (SHiP) experiment~\cite{Bonivento:2013jag,Anelli:2015pba,SHiP:2018yqc,Ahdida:2654870} is a new general purpose fixed target facility proposed at the CERN Super Proton Synchrotron (SPS) accelerator to search for long-lived exotic particles with masses between few hundred MeV and few GeV.
These particles are expected to be predominantly produced in the decays of heavy hadrons.
The facility is therefore designed to maximise the production and detector acceptance of charm and beauty mesons, while providing the cleanest possible environment.
The 400 GeV proton beam extracted from the SPS will be dumped on a high density target with the aim of accumulating $2\times 10^{20}$ protons on target during 5 years of operation.
The charm production at SHiP exceeds that of any existing and planned facility.

\begin{figure}[htb]
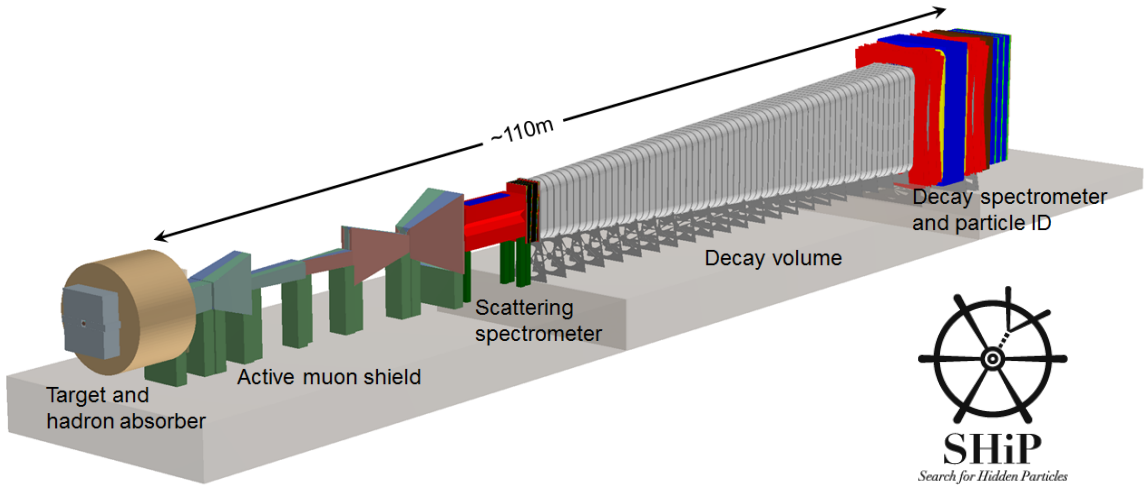

  \begin{center}
    \SHiPOverlay{width=\textwidth}{SHiP_detector2}{80}{0}{width=2cm}
    \caption{Overview of the SHiP experimental facility.}
\label{fig:SHiP_overview}
\end{center}
\end{figure}

A dedicated detector, based on a long vacuum tank followed by a spectrometer and by
particle identification detectors, will allow probing a variety of models with 
light long-lived exotic particles.
Since particles originating in charm and beauty meson decays are produced with a significant transverse momentum with respect to the beam axis, the detector should be placed as close as possible to the target.
A critical component of SHiP is therefore the muon shield~\cite{Akmete:2017bpl}, which deflects  away from the detector the high flux of muons produced in the target, that would otherwise represent a very serious background for hidden particle searches.
To suppress the background from neutrinos interacting in the fiducial volume, the decay volume is maintained under vacuum~\cite{SHiP:2018yqc}.
The detector is designed to reconstruct the exclusive decays of hidden particles and to reduce the background to less than 0.1 events in the sample of $2\times 10^{20}$ protons on target~\cite{Ahdida:2654870}.
The detector consists of a large magnetic spectrometer located downstream of a 50 m long and $5\times 10$ m wide decay volume.
The spectrometer is designed to accurately reconstruct the decay vertex, mass and impact parameter of the decaying particle with respect to the target.
A set of calorimeters followed by muon chambers provide identification of electrons, photons, muons and charged hadrons.
A dedicated timing detector measures the coincidence of the decay products, which allows the rejection of combinatorial background.

The decay volume is surrounded by background taggers to tag neutrino and muon inelastic scattering in the surrounding structures, which may produce long-lived neutral Standard Model  particles, such as $\rm{K}_{\rm{L}}$, that have similar topologies to the expected signal.  

The experimental facility is also ideally suited for studying the interactions of tau neutrinos.
It will therefore host an emulsion cloud chamber based on the Opera concept, upstream of the hidden particle decay volume, followed by a muon spectrometer.
The SHiP facility layout is shown in Fig.~\ref{fig:SHiP_overview}.
Recent progress report~\cite{Ahdida:2654870} outlines the up-to-date experimental design as well as describes changes since the initial technical proposal~\cite{Anelli:2015pba}.

\paragraph{Heavy Neutral Leptons.} Among hypothetical long-lived particles that can be probed by the SHiP experiment are Heavy Neutral Leptons (or HNLs)~\cite{Alekhin:2015byh}.
The idea that HNLs -- also known as right-handed, Majorana or sterile neutrinos -- can be responsible for the smallness of neutrino masses goes back to the 1970s~\cite{Minkowski:1977sc,Yanagida:1979as,Glashow:1979nm,GellMann:1980vs,Mohapatra:1979ia,Mohapatra:1980yp}.
It has subsequently been understood that the same particles could be responsible for the generation of the matter-antimatter asymmetry of the Universe~\cite{Fukugita:2002hu}.
The idea of this scenario, called leptogenesis, was developed since the 1980s (see reviews~\cite{Buchmuller:2004nz,Davidson:2008bu,Shaposhnikov:2009zzb,Pilaftsis:2009pk,Drewes:2017zyw,Chun:2017spz} and references therein).
In particular, it was found that the Majorana mass scale of right-handed neutrinos can be as low as $\mathcal{O}(\text{GeV})$~\cite{Akhmedov:1998qx,Asaka:2005pn,Shaposhnikov:2008pf}, thus providing a possibility for a leptogenesis scenario to be probed at a particle physics laboratory in the near future.

It was demonstrated in 2005 that by adding just three HNLs to the Standard Model one could not only explain  neutrino oscillations and the origin of the baryon asymmetry of the Universe, but also provide a dark matter candidate~\cite{Asaka:2005an,Asaka:2005pn}.
Two of the HNLs should have masses in the GeV range, see~\cite{Boyarsky:2009ix} for a review.
This model, dubbed Neutrino Minimal Standard Model (or $\nu$MSM), is compatible with all the measurements so far performed by accelerator experiments and at the same time provides a solution for the puzzles of modern physics~\cite{Shaposhnikov:2007nj,Boyarsky:2009ix}.
This made models with GeV scale HNLs a subject of intensive theoretical studies in the recent years~\cite{Canetti:2012zc,Drewes:2012ma,Canetti:2012kh,Canetti:2012vf,Garbrecht:2014aga,Shuve:2014zua,Canetti:2014dka,Gago:2015vma,Hambye:2016sby,Hernandez:2016kel,Caputo:2016ojx,Ghiglieri:2016xye,Drewes:2016gmt,Ghiglieri:2017gjz,Caputo:2017pit,Eijima:2017anv,Chun:2017spz,Asaka:2017rdj,Eijima:2017cxr,Antusch:2017pkq,Eijima:2018qke}.

HNLs are massive Majorana particles that possess neutrino-like interactions with $W$ and $Z$ bosons (the interaction with the Higgs boson does not play a role in our analysis and will be ignored).
The interaction strength is suppressed compared to that of ordinary neutrinos by flavour dependent mixing angles $U_\alpha \ll 1$ ($\alpha=\{e,\mu,\tau\}$).
Thus, even the simplest HNL model contains 4 parameters: the HNL mass $M_N$ and 3 mixing angles $U_\alpha^2$.\footnote{The mixing angles $U_\alpha$ are in general complex numbers.
However, the properties of HNLs that are important for us depend only on $|U_\alpha|$.
Throughout this work we will write $U_\alpha^2$ instead of $|U_\alpha|^2$ for compactness.
} The idea of experimental searches for such particles goes back to the 1980s~(see e.g.\cite{Shrock:1981cq,Shrock:1982sc,Shrock:1980ct,Shrock:1981wq,Gronau:1984ct}) and a large number of experiments have searched for them in the past (see review of the past searches in~\cite{Gorbunov:2007ak,Atre:2009rg,Deppisch:2015qwa}).
HNLs are being searched at currently running experiments, including LHCb, CMS, ATLAS, T2K, Belle and NA62~\cite{Aaij:2014aba,Khachatryan:2015gha,Aad:2015xaa,Sirunyan:2018mtv,Izmaylov:2017lkv, Mermod:2017ceo,CortinaGil:2017mqf,Liventsev:2013zz}.

\begin{table}[!t]
  \centering
  \begin{tabular}{|c|c|c|c|c|}
    \hline
    $pN$ cross-section& $\bar c c$ fraction & $\bar b b$ fraction & \multicolumn{2}{c|}{Cascade enhancement $f_{\rm cascade}$} \\
    \cline{4-5}
    $\sigma_{pN}$~\cite{Anelli:2015pba}     & $ X_{\bar c c}$~\cite{Abt:2007zg}       &  $X_{\bar b b}$~\cite{Lourenco:2006vw}      & charm \cite{CERN-SHiP-NOTE-2015-009}                                   & beauty \cite{CERN-SHiP-NOTE-2015-009} \\
    \hline
    $10.7$~mb          & $ 1.7\times 10^{-3}$ & $1.6 \times 10^{-7}$& 2.3                 & 1.7                                           \\
    \hline
  \end{tabular}
  \caption{Charm and beauty production fractions and cascade enhancement factors for the SHiP experiment. Cross-section $\sigma_{pN}$ is an average proton-nucleon inelastic cross-section for the molybdenum target~\cite{Anelli:2015pba}.}
  \label{tab:sigmas}
\end{table}

The sensitivity of the SHiP experiment to HNLs was previously explored for several benchmark models~\cite{Anelli:2015pba,CERN-SHiP-NOTE-2016-003,Gninenko:2013tk} assuming particular ratios between the three HNL mixing angles~\cite{Gorbunov:2007ak}. This paper updates the previous results in a number of important ways.
A recent work~\cite{Bondarenko:2018ptm} revised the branching ratios of HNL production and decay channels.
In addition, the estimates of the numbers of $D$- and $B$-mesons now include cascade production~\cite{CERN-SHiP-NOTE-2015-009}.
We update the lower limit of the SHiP sensitivity region and also evaluate the upper bound for the first time.
We discuss potential impact of HNL production from $B_c$ mesons.
Moreover, our current sensitivity estimates are not limited to a set of benchmark models.
Rather, we compute a \emph{sensitivity matrix} -- a model-independent tool to calculate the SHiP sensitivity for any model of HNL flavour mixings.

The paper is organised as follows.
Section~\ref{sec:background} describes the simulation of HNL events.
The resulting sensitivity curves for mixing with each individual flavour, for the benchmark models of Ref.~\cite{Anelli:2015pba} as well as the sensitivity matrix --  are discussed in Section~\ref{sec:results}.
We present our method to evaluate the SHiP sensitivity to HNLs in a model-independent way in Section~\ref{sec:sensitivity-matrix} and conclude in Section~\ref{sec:conclusion}.

\section{Monte Carlo simulation of heavy neutral leptons at SHiP}
\label{sec:background}

\begin{table}[t]
\centering
    \begin{tabular}{|c|c|}
        \hline
        meson & $f(q\to \text{meson})$ \\
        \hline
        $D^+$ & $0.207$ \\
        \hline
        $D^{0}$ & $0.632$ \\
        \hline
        $D_{s}$ & $0.088$ \\
        \hline
        $\Jpsi$ & $0.01$ \\
        \hline
    \end{tabular}~\begin{tabular}{|c|c|}
        \hline
        meson & $f(q\to \text{meson})$ \\
        \hline
        $B^+$ & $0.417$ \\
        \hline
        $B^{0}$ & $0.418$ \\
        \hline
        $B_{s}$ & $0.113$ \\
        \hline
        $B_{c}$ & $\le 2.6\times 10^{-3}$ \\
        \hline
    \end{tabular}
    \caption{Production fraction and expected number of different mesons in SHiP taking into account cascade production~\cite{Graverini:2133817}. For $f(b\to B_c)$ see text for details.}
    \label{tab:meson_n}
\end{table}

\begin{figure}[!t]
    \centering
    \includegraphics[width=0.49\textwidth]{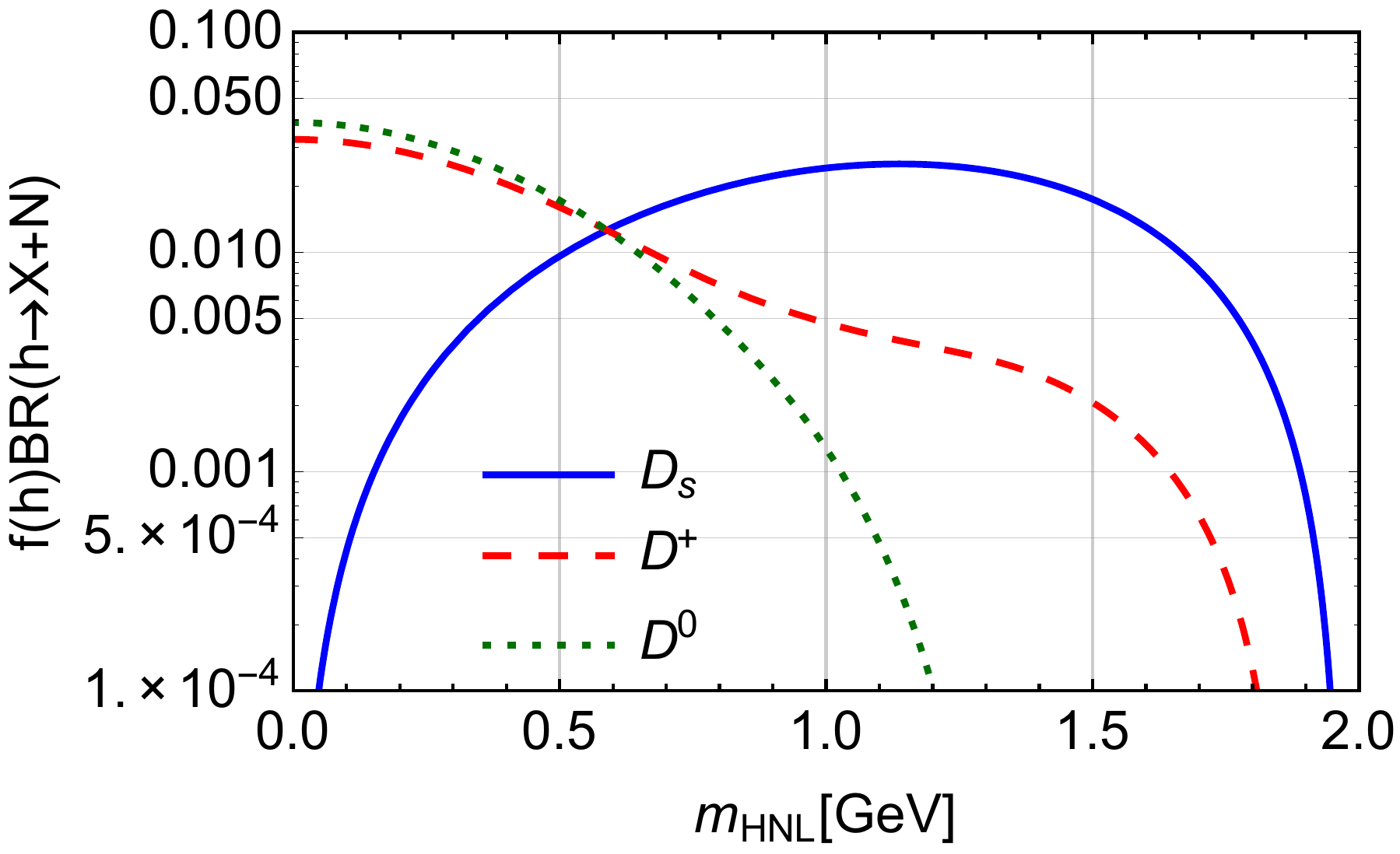}~
    \includegraphics[width=0.46\textwidth]{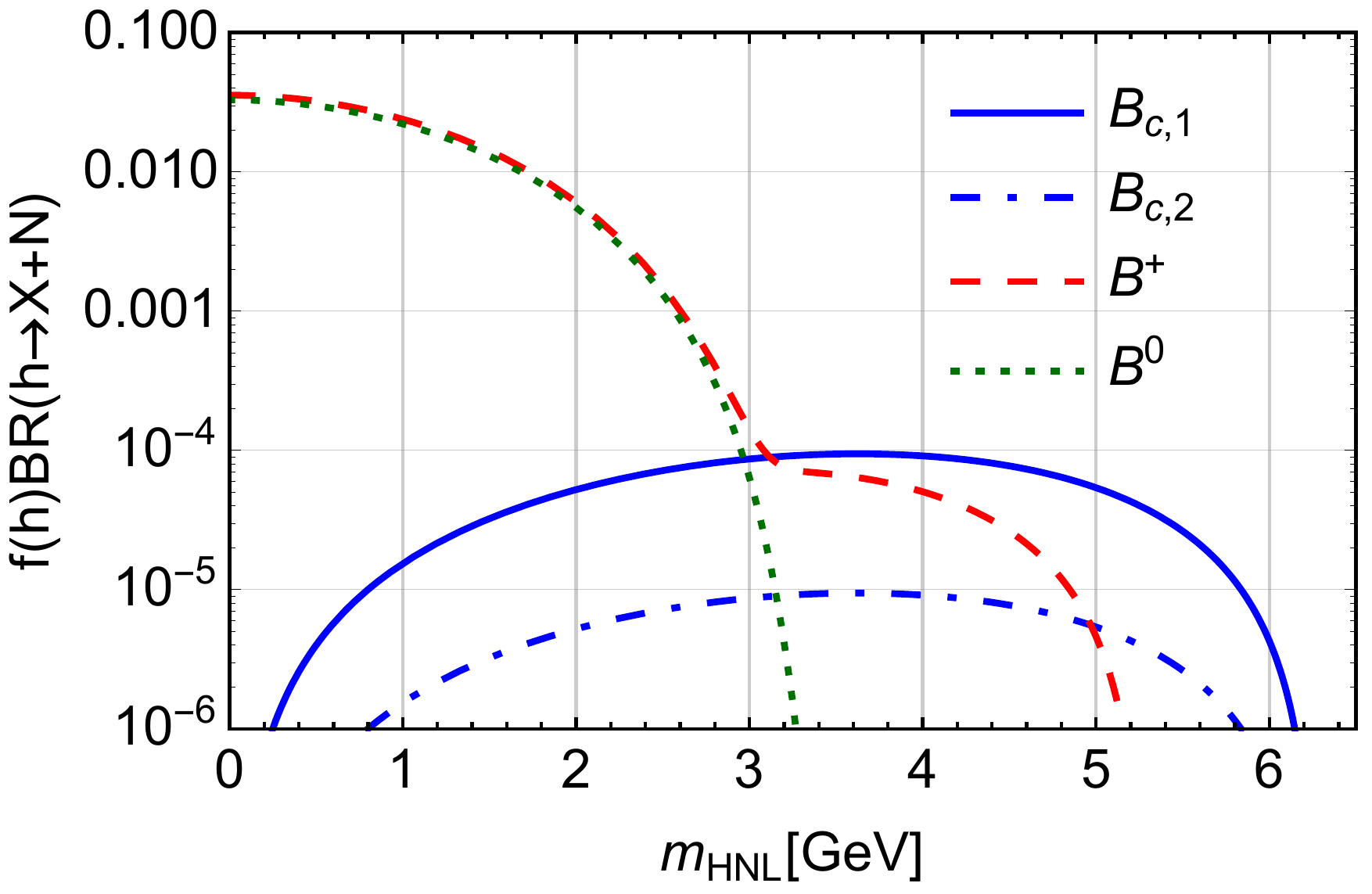}
    \caption{HNL production branching ratios multiplied with the production fraction of the meson decaying into HNL, for charm (left) and beauty (right) mesons~\cite{Bondarenko:2018ptm}. The mixing angles have been set to $U_e^2=1$, $U_\mu^2=U_\tau^2=0$.
The production from $D^+$ and $B^+$ remains relevant for higher masses for $D^0$ and $B^0$ because of the fully leptonic decays $h^+\to N+ \ell^+$.
      The $B_c$ production fraction is unknown (see text for details) and
      we show  two examples: $f(b\to B_c) = 2\times 10^{-3}$ ($B_{c,1}$ line) and $f(b\to B_c) = 2\times 10^{-4}$ ($B_{c,2}$ line).
    }
    \label{fig:ship_production}
  \end{figure}

A detailed Monte Carlo simulation suite for the SHiP experiment, \fship, was developed based on the \texttt{FairRoot} software framework~\cite{FairRoot}.
In \fship simulations primary collisions of protons are generated with \texttt{Pythia 8}~\cite{Sjostrand:2007gs} and the subsequent propagation and interactions  of particles simulated with \texttt{GEANT4}~\cite{Agostinelli:2002hh}.
Neutrino interactions are simulated with \texttt{GENIE}~\cite{Andreopoulos:2009rq}; heavy flavour production and inelastic muon interactions with \texttt{Pythia 6}~\cite{Sjostrand:2006za} and \texttt{GEANT4}.
Secondary heavy flavour production in cascade interactions of hadrons originated by the initial proton collision~\cite{CERN-SHiP-NOTE-2015-009} is also taken into account, which leads to an increase of the overall HNL production fraction (see Table~\ref{tab:sigmas}).
The SHiP detector response is simulated using \texttt{GEANT4}.
The pattern recognition algorithms applied to the hits on the straw spectrometer are described in~\cite{CERN-SHiP-NOTE-2015-002}, and the algorithms for particle identification are presented in~\cite{CERN-SHiP-NOTE-2017-00}.

The simulation takes the HNL mass $M_N$ and its three flavour mixings $U_e^2$, $U_\mu^2$, $U_\tau^2$ as input parameters. For the pure HNLs mixing to a single SM flavour, the number of detected HNL events $N_{\text{events}}$ is estimated as\footnote{The case of the general mixing ratio is discussed in Section~\ref{sec:sensitivity-matrix}.}
\begin{equation}
  \label{eq:3}
    N_{\text{events}} = N_{\text{prod}} \times P_{\text{det}}
\end{equation}
where $N_{\text{prod}}$ is the number of produced HNLs that fly in the direction of the fiducial volume and $P_{\text{det}}$ is the probability of HNL detection in the Hidden Sector detector.
The number of produced HNLs is
\begin{equation}
  \label{eq:4}
  N_{\text{prod}} = \sum_{q \in (c,b)}
  N_{q}\times \sum_{h} f(q\to h) \times \text{BR}(h\to N + X)\times \epsilon_{\text{decay}},
\end{equation}
where $f(q\to h)$ is the $h$ meson  production fraction\footnote{The meson production fraction is the probability that a quark of a given flavour hadronizes into the given meson. In the sum over hadrons we consider only lightest hadrons of a given flavour that have only weak decays.
  Higher resonances have negligible branching to HNLs as they mostly decay via strong interactions.} at SHiP (see Table~\ref{tab:meson_n}), $\text{BR}(h\to N + X)$ is the mass dependent inclusive branching ratios for $h$ mesons decays with HNL in the final state and $ \epsilon_{\text{decay}}$ is the \emph{geometrical acceptance} -- the fraction of produced HNLs that fly into direction of the fiducial volume.
Fig.~\ref{fig:ship_production} shows the product between the meson production fraction and its inclusive decay branching fraction into sterile neutrinos. Finally,
$N_q$ is the total number of produced quarks and antiquarks of the given flavour $q$ taking into account the quark-antiquark production fraction $X_{\bar q q}$ and the cascade enhancement factor $f_{\rm cascade}$ given in Table~\ref{tab:sigmas}, 
\begin{equation}
 N_q = 2\times X_{\bar q q} \times f_{\rm cascade} \times N_{\rm POT}.
\end{equation}

The HNL \emph{detection probability} is given by
\begin{equation}
  \label{eq:matrix:2}
  P_{\text{det}} = P_{\text{decay}}\times \text{BR}(N\to\text{visible}) \times \epsilon_{\det},
\end{equation}
where $\text{BR}(N\to\text{visible})$ is the total HNL decay branching ratio into visible channels (see HNL decay channels in Appendix~\ref{sec:hnl-decays}), $P_{\text{decay}}$ is the probability that the HNL decays inside the fiducial volume,
\begin{equation}
  P_{\text{decay}} = \exp\left(-\frac{l_{\text{ini}}}{l_{\text{decay}}}\right) - \exp\left(-\frac{l_{\text{fin}}}{l_{\text{decay}}}\right),
  \label{eq:Pdecay1}
\end{equation}
where $l_{\text{ini}}$ is the distance travelled by HNL before it entered the decay vessel;
$l_{\text{fin}}$ is the distance to the end of the decay vessel along the HNL trajectory;
$l_{\text{decay}} = c \gamma \tau_N$ is the HNL decay length ($\gamma$ and $\tau_N$ being HNL gamma factor and proper lifetime).
Finally, $\epsilon_{\text{det}}$ is the efficiency  of detecting the charged daughters of the decaying HNL. It takes into account the track reconstruction efficiency and the selection efficiency, further described in~\cite{Anelli:2015pba,CERN-SHiP-NOTE-2017-00,CERN-SHiP-NOTE-2016-003}. In order to distinguish the signal candidates from  possible SM background,  we put a criteria that at least two charged tracks reconstructed to the decay point are present.
The reconstruction efficiencies for the decay channels $N \to \mu\mu\nu$ and $N\to \mu\pi$ are given in e.g.\ \cite[Section 5.2.2.2]{Anelli:2015pba}.
Using \texttt{FairShip}, a scan was done over the HNL parameter space. For each set of HNL parameters we ran a simulation with 300 HNL events, produced randomly from decay of mesons. We determined $P_\decay$, $\epsilon_\decay$ and $\epsilon_{\text{det}}$ in each of them and average over simulations to find the expected number of detected events, $\bar{N}_{\text{events}}$.

For HNLs with masses $M_N \lesssim 500$~MeV kaon decays are the dominant production channel.
While $\mathcal{O}(10^{20})$ kaons are expected at SHiP, most of them are stopped in the target or hadron stopper before decaying.
As a consequence, only HNLs originating from charm and beauty mesons are included in the estimation of the sensitivity. SHiP can however explore the $\nu$MSM parameter space down to the constraints given by Big Bang nucleosynthesis observations~\cite{Dolgov:2000jw,Ruchayskiy:2012si}, even with this conservative assumption.
It is expected that the NA62 experiment will also probe the region below the kaon mass~\cite{Dobrich:2017yoq}.

For HNL masses  $M_N \gtrsim 3$~GeV the contribution of $B_c$ mesons to the HNL production
can be relevant because the $B_c^+ \to N + \ell^+$ decay width is proportional to  the CKM matrix element $|V_{cb}|^2$, while the decays of $B^+$ are proportional to $|V_{ub}|^2$~\cite{Gorbunov:2007ak,Bondarenko:2018ptm}. The ratio $|V_{cb}|^2/|V_{ub}|^2 \sim 10^2$, which explains the relative importance of $B_c$ channels even for small production fraction $f(b\to B_c)$. 
This production fraction has not been measured at the SHiP center of mass energy.
If the $B_c$ production fraction at SHiP is at the LHC level,  its contribution will be dominant.
However, at some unknown energy close to the $B_c$ mass this production fraction becomes negligible.
The existing Tevatron measurement place $f(b\to B_c) = 2.08^{+1.06}_{-0.95}\times 10^{-3}$ at $\sqrt{s}=1.8$~TeV~\cite{Cheung:1999ir}. 
More recent LHCb measurement at $\sqrt{s}=7$ and $8$~TeV gave $f(b\to B_c)/f(b\to B^+) = 0.008 \pm 0.004$~\cite{Aaij:2017kea}. 
Using $f(b\to B^+)=0.33$ from the LHCb measurement performed at $\sqrt{s} = 7$~TeV~\cite{Aaij:2011jp}, one obtains $f(b\to B_c) = 2.6\times 10^{-3}$.
Theoretical evaluations have mostly been performed for TeV energies (see e.g.~\cite{Berezhnoy:1995au,Berezhnoy:2004gc,Chang:2005bf,Berezhnoy:2010wa}) with the exception of the works~\cite{Kolodziej:1995nv,Kolodziej:1997um} that computed the production fraction down to energies of tens of GeV (where they found the fraction to be negligible).  
However, by comparing predictions of~\cite{Kolodziej:1997um} with LHCb or Tevatron measurements, we see that \textit{(i)} it underpredicts the value of $f(b\to B_c)$ by about an order of magnitude at these energies and \textit{(ii)} it predicts stronger than observed change of the production fraction between LHC and Tevatron energies.
Therefore we have to treat $f(b\to B_c)$ as an unknown parameter somewhere between its LHC value and zero
and provide two estimates: an optimistic estimate for which $f(b\to B_c)$ is at the LHC level and a pessimistic estimate where we do not include $B_c$ mesons at all. In the simulation we take the angular distribution of $B_c$ mesons to be the same as that of $B^+$ mesons, based on comparisons performed with the BCVEGPY~\cite{Chang:2003cq} and FONLL~\cite{Cacciari:1998it,Cacciari:2001td} packages, while we rescale the energy distribution according to the meson mass.

Detailed background studies have proven that the yield of background events passing the online and offline event selections is negligible~\cite{Anelli:2015pba}.
Therefore, the 90\% confidence region is defined as the region of the parameter space where one expects on average $\bar{N}_{\text{events}} \ge 2.3$ reconstructed HNL events, corresponding to the discovery threshold with an expected background yield of 0.1 events.

\section{SHiP sensitivity for benchmark HNL models}
\label{sec:results}

\begin{figure}[!t]
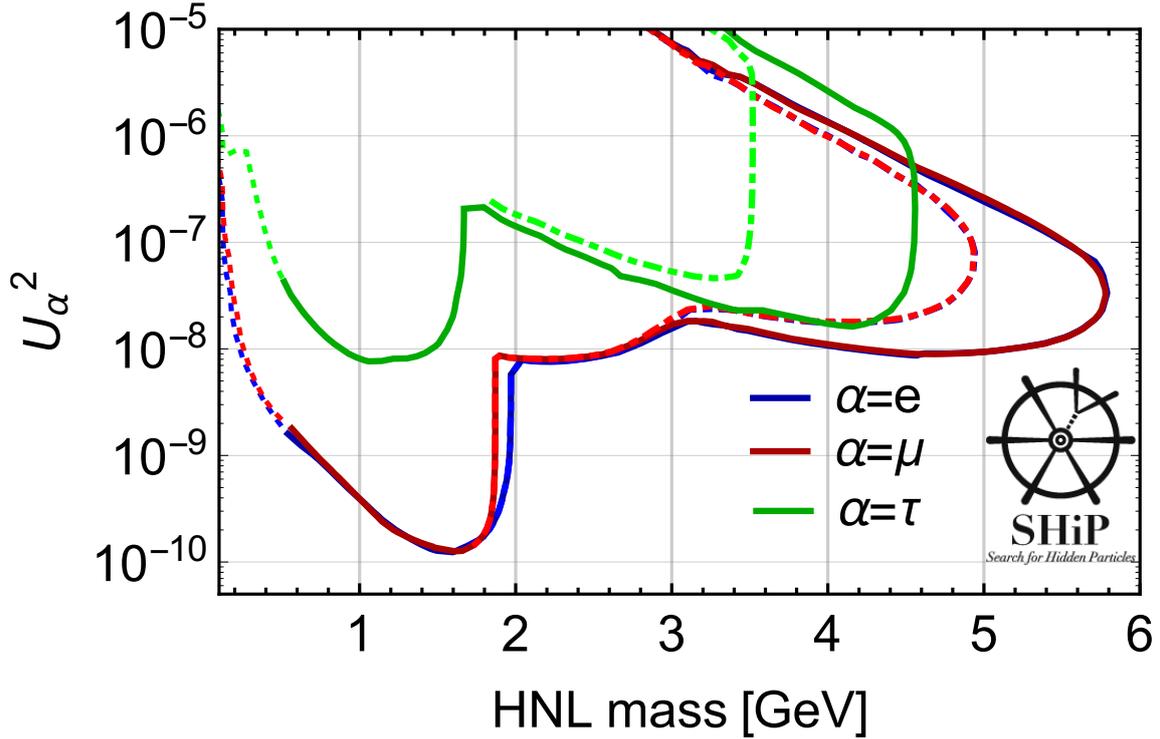

  \centering
  \SHiPOverlay{width=\textwidth}{SHiP_90CL}{85}{15}{width=2cm}
  \caption[SHiP sensitivity to individual flavours]{SHiP sensitivity curves (90\% CL) for HNLs mixing to a single SM flavour: electron (blue), muon (red) and tau (green).
    To indicate the uncertainty related to the unknown production fraction of $B_c$ meson (see text for details), we show two types of curve for each flavour.
    Solid curves show the sensitivity contours when the production fraction of $B_c$ mesons equals to that at LHC energies: $f(b \to B_c)= 2.6\times 10^{-3}$.
    Dashed-dotted lines do not include contributions from $B_c$.
    Below 0.5~GeV only production from $D$ and $B$ mesons is included (dotted lines).
  }
  \label{fig:one_flavour}
\end{figure}
Figure~\ref{fig:one_flavour} presents the 90\% C.L. sensitivity curves for HNLs mixing to only one SM flavour.
The sensitivity curves have a characteristic ``cigar-like shape'' for masses $M_N > 2$~GeV.  The \emph{upper boundary} is
determined by the condition that the decay length of a produced particle
becomes comparable with the distance between the target and the decay volume,
and therefore the HNLs produced at the target may not reach the decay volume, see Eq.~\eqref{eq:Pdecay1}.
For masses $M_N < 2$~GeV such an upper boundary also exists, but it is outside the plot range, owing to a much larger number of parent $D$ mesons.
The \emph{lower boundary} of the
sensitivity region is determined by the parameters at which decays become too
rare (decay length much larger than the detector size).  The intersection of the upper and lower boundaries defines the \emph{maximal mass} which can be probed at the experiment.

\begin{figure}[!t]
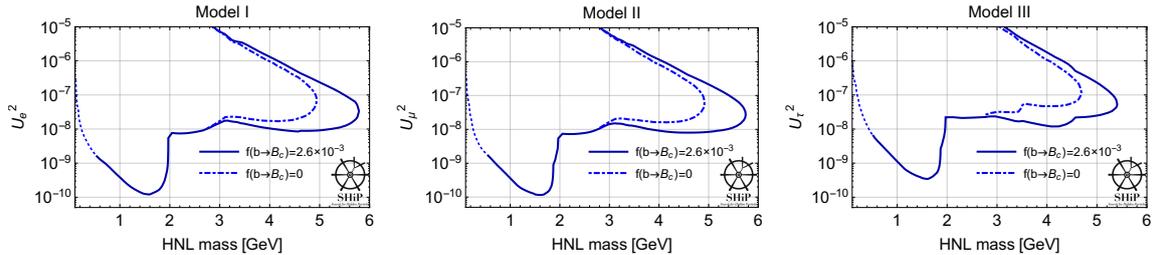

  \centering
  \SHiPOverlay{width=0.32\textwidth}{SHiP_BC6_90CL}{88}{13}{width=5mm}\SHiPOverlay{width=0.32\textwidth}{SHiP_BC7_90CL}{88}{13}{width=5mm}\SHiPOverlay{width=0.32\textwidth}{SHiP_BC8_90CL}{88}{13}{width=5mm}
  \caption[Sensitivity for benchmark models]{Sensitivity curves for 3 benchmark models I--III ($90\%$CL). Individual curves are explained in Fig.~\ref{fig:one_flavour}.}
  \label{fig:benchmark}
\end{figure}

We also provide updated sensitivity estimates for the three benchmark models I--III presented in the Technical Proposal~\cite{Anelli:2015pba,CERN-SHiP-NOTE-2016-003}.
These models allow to explain neutrino flavour oscillations while at the same time maximizing the mixing to one particular flavour, and are defined by the following ratios of flavour couplings~\cite{Gorbunov:2007ak}:
\begin{compactitem}[--]
\item I. $U_e^2 : U_\mu^2 : U_\tau^2 = 52 : 1 : 1$ 
\item II. $U_e^2 : U_\mu^2 : U_\tau^2 = 1 : 16 : 3.8$ 
\item III. $U_e^2 : U_\mu^2 : U_\tau^2 = 0.061 : 1 : 4.3$ 
\end{compactitem}
The sensitivity curves for these models are shown in Fig.~\ref{fig:benchmark}.

\section{Model independent SHiP sensitivity}
\label{sec:sensitivity-matrix}

In this Section we provide an efficient way to estimate the SHiP sensitivity to an HNL model with an arbitrary ratio $U_e^2 : U_\mu^2 : U_\tau^2$.
It is based on the observation that the dependence of the number of events, $N_\event$, on the mass and  mixing angles of HNL factorizes, and therefore all relevant information can be extracted from a handful of simulations, rather than from a scan over an entire 4-dimensional HNL parameter space $(M_N, U_e^2 , U_\mu^2 , U_\tau^2)$.

All information about the HNL production in a particular experiment is contained in $N_\alpha(M_N)$ -- the number of HNLs that would be produced through all possible channels with the mixings $U^2_\alpha = 1$ and $U^2_{\beta \neq \alpha} = 0$:
\begin{equation}
  \label{eq:7}
  N_\alpha \equiv \sum_{\text{hadrons }h} N_h \sum_{\text{channels}} \BR(h {\to} N + X_\alpha)\epsilon_{\text{decay},\alpha}\Bigr|_{U^2_\alpha = 1; U^2_{\beta \neq \alpha} = 0}
\end{equation}
Here $N_h$ is the number of hadrons of a given type $h$, $ \BR(h {\to} N + X_\alpha)$ is the branching ratio for their decay into an HNL plus any number of other particles $X_\alpha$ with total lepton flavour number $L_\alpha =1$ and $\epsilon_{\text{decay},\alpha}$ is the geometrical acceptance of HNL that in general depends not only on the mass but also on the flavour. 
The overall number of HNLs (given by Eq.~\eqref{eq:4}) produced via the mixing with the flavour $\alpha$ and flying in the direction of the decay vessel is given by
{\begin{equation}
  \label{eq:5}
  N_{\pr,\alpha}(M_N|\vecU) = U_\alpha^2 N_\alpha(M_N).
\end{equation}}

The decay probability $P_\decay$ should be treated differently, depending on the ratio of the decay length and the distance from the target to the decay vessel. It also depends on the production channel through the mean gamma factor $\gamma_\alpha$ entering  the decay length.

In the limit when the decay length much larger than the distance between the beam target and the exit lid of the SHiP decay volume, the {$U^2_\beta$} dependence of the decay probability can be accounted for similarly to Eq.~\eqref{eq:5}:
%
\begin{equation}
\label{eq:Plinear}
    P_{\decay,\alpha}^{\text{linear}}(M_N|\vecU) = \frac{l_{\text{fin}} - l_{\text{ini}}}{\gamma_{\alpha} c\hbar} \sum_\beta U_\beta^2 \Gamma_{\beta}(M_N),
\end{equation}
where $\Gamma_{\beta}$ is a decay width of the HNL of mass $M_N$ that has mixing angles $U_\beta^2 = 1$, $U_{\alpha\ne\beta}^2 = 0$, the definitions of lengths $l_{\text{ini}},l_{\text{fin}}$ are given after Eq.~\eqref{eq:Pdecay1}. The index $\alpha$ in Eq.~\eqref{eq:Plinear} indicates that the HNL was produced via mixing $U_\alpha^2$ (although can decay through the mixing with any flavour), so $\gamma_{\alpha}$ is the mean gamma factor of HNLs produced through the mixing with the flavour $\alpha$.

In the general case, when the decay length $l_\decay$ is not necessarily larger than $l_{\text{fin}}$, the analogous decay probability $P_{\text{decay},\alpha}$ can  be expressed via~\eqref{eq:Plinear} as follows:
{\begin{multline}
  P_{\text{decay},\alpha}(M_N|\vecU) = \bigg[\exp\left(-\frac{l_{\text{ini}}}{l_{\text{fin}}-l_{\text{ini}}} P_{\decay,\alpha}^{\text{linear}}(M_N|\vecU)\right) 
  - \\
  \exp\left(-\frac{l_{\text{fin}}}{l_{\text{fin}}-l_{\text{ini}}}P_{\decay,\alpha}^{\text{linear}}(M_N|\vecU) \right)\bigg]\times \text{BR}(N\to\text{visible}),
\end{multline}}
where $\text{BR}(N\to\text{visible})$ is the probability that the HNL decays into the final states detectable by SHiP.

Finally, we define the HNL detection efficiency as
{\begin{equation}
\epsilon_{\det}(M_N|\vecU) = \sum_{\beta} \text{BR}(N\to X_{\beta}) \times \epsilon_{\det,\beta},
\end{equation}}
where {$\text{BR}(N\to X_{\beta})$} is the branching ratio of a decay through the mixing angle {$\beta$} and {$\epsilon_{\det,\beta}$} is the probability that the HNL decay products are successfully detected. 

As a result, the number of detected events is given by
{\begin{equation}
  \label{eq:1}
  N_\decay\Bigl( M_N\bigl| \vecU\Bigr) = 
  \sum_\alpha N_{\pr,\alpha}(M_N|\vecU) P_{\text{decay},\alpha}(M_N|\vecU) \epsilon_{\det}(M_N|\vecU).
\end{equation}}
\emph{We see that it is sufficient to know 9 functions of the HNL mass -- $N_\alpha(M_N)$, $P_{\text{decay},\alpha}^{\text{linear}}(M_N)$ and $\epsilon_{\det,\alpha}(M_N)$ -- to determine the number of detected events for any combination of the mixing angles.}

To determine these numbers we ran $9$ Monte Carlo simulations for each mass.
  We first ran 3 simulations with vectors $\vecU = (x,0,0)$, $\vecU = (0,x,0)$, $\vecU = (0,0,x)$, where $x$ is any sufficiently small number such that $l_\decay \gg l_{\det}$.
  We then ran a set of 6 non-physical simulations, where a particle is produced solely via channel $\alpha$ and decays solely through the channel $\beta \neq \alpha$.
  {Using results of these simulations we extract $N_{\alpha}$, $P_{\alpha}$ and $\epsilon_{\det,\alpha}$ values} that allow us to generate the expected number of detected events for any values of masses and couplings.

  The results are available at Zenodo platform~\cite{zenodo} with instructions for reading the file and generating sensitivity curves at different confidence levels.

\section{Conclusion}
\label{sec:conclusion}

\begin{figure}[!t]
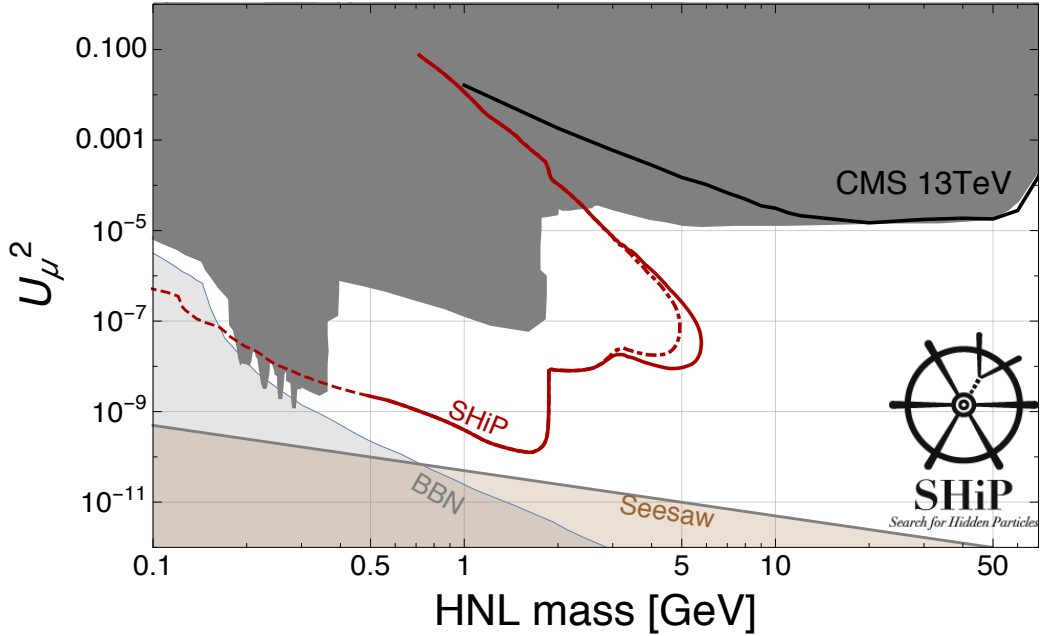

 \centering
  \SHiPOverlay{width=0.9\textwidth}{Umu_previous_no_future_experiments}{85}{10}{width=2cm}
  \caption{Parameter space of HNLs and potential reach of the SHiP experiment for the mixing with muon flavour.
    Dark gray area is excluded from previous experiments, see e.g.~\cite{Alekhin:2015byh}.
    Black solid line is the recent bound from the CMS 13~TeV run~\cite{Sirunyan:2018mtv}.
    Solid and dashed-dotted red lines indicate the uncertainty, related to the production fraction of $B_c$ mesons at SHiP energies that has not been measured experimentally or reliably calculated (see Section~\ref{sec:background} for details).
    The sensitivity of SHiP below kaon mass (dashed line) is based on the number of HNLs produced in the decay of $D$-mesons only and does not take into account HNL production from kaon decays.
    The primordial nucleosynthesis bounds on HNL lifetime are from~\cite{Dolgov:2000jw}.
    The seesaw line indicates the parameters obeying the seesaw relation $|U_\mu|^2\sim m_\nu/M_N$, where for active neutrino mass we substitute $m_\nu = \sqrt{\Delta m_{\text{atm}}^2} \approx \unit[0.05]{eV}$~\cite{Alekhin:2015byh}.}
  \label{fig:HNLbounds}
\end{figure}

Using a detailed Monte Carlo simulation of HNL production in decays of charm and beauty mesons, and of the detector response to the signal generated by a decaying HNL, we calculated the sensitivity of the SHiP experiment to HNLs, updating the results presented in the Technical Proposal~\cite{Anelli:2015pba}. In particular, we assess the potential impact of HNL production from $B_c$ mesons decay, showing its influence on the extent of the probed HNL mass range. We take into account cascade production of $B$ and $D$ mesons as well as revised estimates of branching ratios of HNL production and decay, and we extend our calculation to masses below $\sim 500$~MeV, where SHiP has a potential to fully explore the allowed region. Finally, we present our results as a publicly available dataset, providing a model-independent way to calculate the SHiP sensitivity for any pattern of HNL flavour mixings.

The SHiP experiment offers an increase of up to 3 orders of magnitude in the sensitivity to heavy neutral leptons, Fig.~\ref{fig:HNLbounds}. It is capable of probing cosmologicaly interesting region of the HNL parameter space, and of potentially discovering the origin of neutrino masses and of the matter-antimatter asymmetry of the Universe.

\section*{Acknowledgments}

The SHiP Collaboration wishes to thank the Castaldo company (Naples, Italy) for their contribution to the development studies of the decay vessel. The support from the National Research Foundation of Korea with grant numbers of 2018R1A2B2007757, 2018R1D1A3B07050649, 2018R1D1A1B07050701,
2017R1D1A1B03036042, \\ \noindent 2017R1A6A3A01075752, 2016R1A2B4012302, and 2016R1A6A3A11930680 is acknowledged.
The support from the European Research Council (ERC) under the European Union’s Horizon 2020 research and innovation programme (GA No 694896) is acknowledged. 
The support from the Russian Foundation for Basic Research (RFBR) and the support from the TAEK of Turkey are acknowledged.

\appendix

\section{HNL decays}
\label{sec:hnl-decays}

For completeness we list the relevant HNL decay channels in Table~\ref{tab:decaychannels} (reproduced from~\cite{Bondarenko:2018ptm}).
\begin{small}
\LTcapwidth=\textwidth
\begin{longtable}{|c|c|c|c|c|c|}
  \caption[Table of the relevant HNL decay channels]{\small
  List of the relevant HNL decay channels with branching
  ratio above $1\%$ covering the HNL mass range up to $5$~GeV implemented in \fship.
  The
  numbers are provided for 
  $|U_e|^2=|U_\mu|^2=|U_\tau|^2$.
  For neutral current channels (with neutrinos in the final state) the sum over
  neutrino flavours is taken, otherwise the lepton flavour is shown explicitly.\newline
  Columns: (1) the HNL decay channel.
  (2)
  The HNL mass at which the channel opens.
  (3) The HNL mass
  starting from which the channel becomes relevant (branching ratio of this channel exceeds $1\%$).
  For multimeson
  final states we provide our best-guess estimates.
  (4) HNL mass above which the
  channel contributes less than 1\%, with ``---'' indicating that the channel is
  still relevant at $M_N=5$~GeV.
  (5)
  The maximum branching ratio of the channel for $M_N<5$~GeV.
  (6) Reference to the appropriate formula for decay width in ref.~\cite{Bondarenko:2018ptm}.
\label{tab:decaychannels}} \\
\hline
  Channel & Opens at& Relevant from & Relevant up to & Max BR & Reference \\
  & [MeV]  &  [MeV]  &  [MeV] &  [\%] & in~\cite{Bondarenko:2018ptm}\\
  \endfirsthead 
\caption[]{(continued)} \\
\hline
 Channel & Opens at& Relevant from & Relevant up to & Max BR & Reference \\
  & [MeV]  &  [MeV]  &  [MeV] &  [\%] & in~\cite{Bondarenko:2018ptm}\\                                                                     \endhead \hline
$N\to \nu_{\alpha} \nu_{\beta} \bar{\nu}_{\beta}$ 
	& $\sum m_{\nu}\approx 0$ & $\sum m_{\nu}\approx 0$ & --- & 100 & (3.5) \\ \hline
$N\to \nu_\alpha e^+e^-$ 
	& 1.02 & 1.29 & --- & 21.8 & (3.4) \\ \hline
$N\to \nu_\alpha \pi^0$ 
	& 135 & 136 & 3630 & 57.3 & (3.7) \\ \hline
$N\to e^- \pi^+$ 
	& 140 & 141 & 3000 & 33.5 & (3.6) \\ \hline
$N\to \mu^-\pi^+$ 
	& 245 & 246 & 3000 & 19.7 & (3.6) \\ \hline
$N\to e^- \nu_\mu\mu^+$ 
	& 106 & 315 & --- & 5.15 & (3.1) \\ \hline
$N\to \mu^- \nu_e e^+$ 
	& 106 & 315 & --- & 5.15 & (3.1) \\ \hline
$N\to \nu_\alpha \mu^+ \mu^-$ 
	& 211 & 441 & --- & 4.21 & (3.4) \\ \hline
$N\to \nu_{\alpha} \eta$ 
	& 548 & 641 & 2330 & 3.50 & (3.7) \\ \hline
$N\to e^-\rho^+$ 
	& 770 & 780 & 4550 & 10.4 & (3.8) \\ \hline
$N\to \nu_{\alpha}\rho^0$ 
	& 770 & 780 & 3300 & 4.81 & (3.9) \\ \hline
$N\to \mu^-\rho^+$ 
	& 875 & 885 & 4600 & 10.2 & (3.8) \\ \hline
$N\to \nu_{\alpha} \omega$ 
	& 783 & 997 & 1730 & 1.40 & (3.9)
	\\ \hline
$N\to \nu_{\alpha} \eta'$ 
	& 958 & 1290 & 2400 & 1.86 & (3.7) \\ \hline
$N\to \nu_{\alpha} \phi$ 
	& 1019 & 1100 & 4270 & 5.90 & (3.9)
	\\ \hline
$N\to e^- D_s^{*+}$ 
	& 2110 & 2350 & --- & 3.05 & (3.8) \\ \hline
$N\to \mu^- D_s^{*+}$ 
	& 2220 & 2370 & --- & 3.03 & (3.8) \\ \hline
$N\to e^- D_s^+$ 
	& 1970 & 2660 & 4180 & 1.23 & (3.6) \\ \hline
$N\to \mu^- D_s^+$ 
	& 2070 & 2680 & 4170 & 1.22 & (3.6) \\ \hline
$N\to \nu_{\alpha} \eta_c$ 
	& 2980 & 3940 & --- & 1.26 & (3.7) \\ \hline
$N\to \tau^- \nu_e e^+$
	& 1780 & 3980 & --- & 1.52 & (3.1) \\ \hline
$N\to e^- \nu_\tau \tau^+$
	& 1780 & 3980 & --- & 1.52 & (3.1) \\ \hline
$N\to \tau^- \nu_\mu \mu^+$ 
	& 1880 & 4000 & --- & 1.51 & (3.1) \\ \hline
$N\to \mu^- \nu_\tau \tau^+$ 
	& 1880 & 4000 & --- & 1.51 & (3.1) \\ \hline
\end{longtable}
\end{small}

\bibliographystyle{JHEP} 
\bibliography{ship}

\end{document}